# Increasing the ν = 5/2 gap energy: An analysis of MBE growth parameters

C Reichl[1], J Chen[2], S. Baer[1], C. Rössler[1], T. Ihn[1], K. Ensslin[1], W Dietsche[1,2] and W Wegscheider[1]

[1]Solid State Physics Laboratory, ETH Zurich, CH-8093 Zurich, Switzerland
[2]Max-Planck-Institute for Solid State Research, D-70569 Stuttgart, Germany
Email: creichl@phys.ethz.ch

**Abstract.** The fractional quantized Hall state (FQHS) at the filling factor ν = 5/2 is of special interest due to its possible application for quantum computing. Here we report on the optimization of growth parameters that allowed us to produce two-dimensional electron gases (2DEGs) with a 5/2 gap energy up to 135 mK. We concentrated on optimizing the MBE growth to provide high 5/2 gap energies in "as-grown" samples, without the need to enhance the 2DEG´s properties by illumination or gating techniques. Our findings allow us to analyse the impact of doping in narrow quantum wells with respect to conventional DX-doping in $Al_xGa_{1-x}As$. The impact of the setback distance between doping layer and 2DEG was investigated as well. Additionally, we found a considerable increase in gap energy by reducing the amount of background impurities. To this end growth techniques like temperature reductions for substrate and effusion cells and the reduction of the Al mole fraction in the 2DEG region were applied.

## 1. Introduction

Even 26 years after its discovery in 1987 [1] the first even denominator fractional quantum Hall state (FQHS) is still a hot topic in semiconductor physics (for a detailed recapitulation see e.g. [2, 3]). Due to the proposed non-Abelian behaviour of the associated quasi-particle statistics [4, 5], the 5/2 state is of special relevance for topological quantum computing [6]. However, due to its fragility and at the same time the promise to overcome the decoherence problem in quantum computing by topological protection this state remains enigmatic. Consequently there is great interest for semiconductor-devices featuring 2DEGs sporting a pronounced and stable 5/2 state. Molecular beam epitaxy in the GaAs/AlGaAs material system represents the instrument of choice for creating such systems of ultra-high purity, which are often characterized via the 2DEG´s electron mobility. For more than two decades, a number of high-mobility 2DEGs with 5/2 activation energies up to 560 mK have been analysed [7-21]. However, most of these results required prior illumination of the intrinsic 2DEG [7-9, 11-14, 18] or the use of gating techniques to enhance the quality of or create the 2DEG [16, 19].

As many experiments forbid such preparations (e.g. top-gated 2DEGs are not well suited for illumination due to the nontransparent metal gates and the typically occurring gate leakage currents)

[22-25] there is an intrinsic demand for high-quality 2DEG-structures showing large 5/2 gap energies in "as-grown" condition [e.g. 26].

A lot of effort has been put into identifying the processes that determine a 2DEGs overall quality and with that the reachable 5/2 gap energy [19, 27-31]. There is general consensus that there are two main scattering processes limiting the quality of state-of-the-art 2DEG samples: Background impurity (BI) and remote impurity (RI) scattering. It was found that – given a low level of BI impurities – RI-scattering is the crucial process controlling the 5/2 gap energy [19, 21, 31]. We report on our findings to optimize sample quality and with that the size of the 5/2 activation gap by optimizing the MBE growth itself.

## 2. Experiment

The samples investigated in this work consist of 30 nm wide GaAs/$Al_xGa_{x-1}As$ quantum wells hosting the 2DEG, which are modulation-doped from both sides (see figure 1(a) for a schematic drawing). While the 5/2 state was discovered using structures doped only from the top side [1], the concept of adding a second, "inverted" doping layer below the 2DEG led to a substantial increase not only in electron density and mobility [32], but also in gap energy. Growth rates were about one µm per hour (for GaAs) and the growth temperature was held at 630°C (except for sample F, which was grown at 650°C). For the doping regions the temperature was reduced to 500°C in order to suppress dopant segregation. A variety of growth parameters was changed to investigate their effect on gap energy; see table A for a comprehensive list of relevant growth parameters.

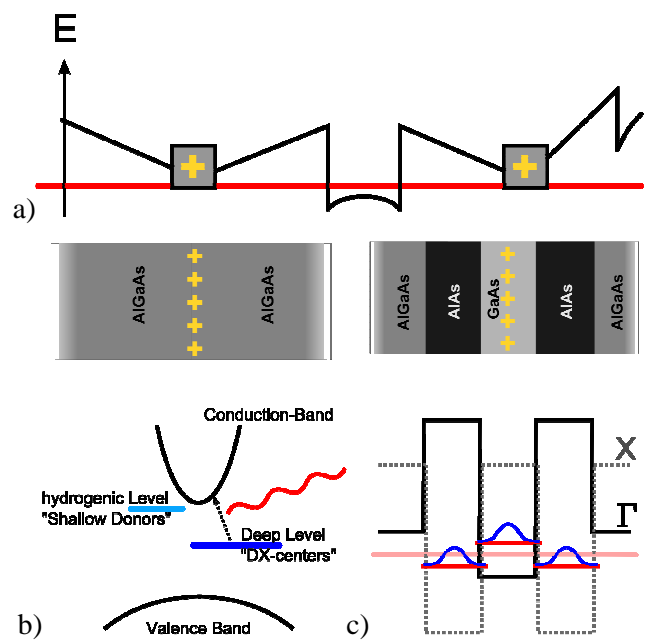

**Figure 1.** (a) Schematic conduction band diagram of the quantum well region of the samples investigated in this work. The Fermi level is indicated in red.
(b) DX-doping scheme: Dopants located in bulk $Al_xGa_{1-x}As$ material are activated by illumination, forming a partially conducting screening layer.
(c) QW-doping scheme: AlAs barriers provide heavy X-band electron states, leading to a non-conducting screening layer.

The electron density of the 2DEG structures was calculated using a self-consistent Schroedinger-Poisson-solver [33], taking into account a change of the effective setback distance $d_{eff}$ by dopant segregation. Comparison with experimental data reveals that the dopant peak shifts by 2.5 nm in growth direction (as opposed to about 4 nm when doping at normal growth temperatures [34]). In order to compensate for this effect the lower setback ($d_l$) was grown 5nm thicker than the upper one ($d_u$), leading to a symmetric set of effective setbacks (eg $d_l$: 75 nm, $d_u$: 70nm; $d_{eff}$: 72.5 nm). The MBE system used to synthesise the samples was optimised for ultra-pure GaAs based semiconductor growth employing a modified Varian Gen II setup. As the UHV quality during growth operations suffers due to the resistive heating of substrate and material sources, we reduced the growth temperature as well as the temperature of the Aluminium source (in the 2DEG region) to minimize the effects of BI-impurities incorporated into the structures. Especially the latter led to a substantial increase in electron mobility (and the gap energy $\Delta_{5/2}$). Aluminium is known to be highly reactive and thus getters background impurities. These impurities are especially effective scatterers when they are incorporated near the quantum well. Therefore a reduction of the aluminium fraction is expected to substantially reduce BI scattering: on the one hand by suppressing the incorporation of background impurities, on the other hand by reducing the amount of aluminium that segregates into the quantum well.

As mentioned above, long range scattering caused by irregular Coulomb potentials emanating from statistically distributed charged donors (RI-scattering) is detrimental for the activation energy of the $\nu=5/2$ FQHS. After the concept of QW-doping being first introduced by Baba et al. [35], it was proposed as a method of screening the 2DEG from such RI-scattering by Friedland et al. [29] and later refined by Umansky et al. However, these samples required massive over-doping of at least $\gamma = 2.5$ in order to acquire fully developed FQHS-minima in the $R_{XX}$-measurements [31]. Here, $\gamma = N_D/N_{min}$ denotes the ratio between actual doping $N_D$ and the minimal doping $N_{min}$ needed in an otherwise identical reference sample to acquire the same electron density. If not stated otherwise (sample I), $\gamma$ is the same for both doping layers. Such an excessive amount of dopants not only causes hysteresis in gated structures (e.g. gate-defined quantum point contacts [36]) but also requires high gate voltages to deplete the doping region before any tuning of the 2DEG is possible. In our samples we confined $\gamma$ to a lower value of about 1.5, providing improved tuning capabilities and reducing the conductance fluctuation that accompany high gate voltages.

Two sets of structures were grown to compare conventional DX-doping – consisting of a δ-layer of dopants in $Al_xGa_{x-1}As$ (see figure 1(b)) – with QW-doping. The latter consists of a δ-layer of dopants in a narrow GaAs quantum well with AlAs barriers to provide electron states in the AlAs X-band with a high effective mass compared to the GaAs Γ-states (figure 1(c)). These X-electrons act as an efficient screening layer to the potential fluctuations. For our samples the width of the GaAs quantum well is 1.4 nm, AlAs layers are 1.9 nm thick. The two sets differ primarily in the setback distance, which allows us to deduce the positive effect of moving the potential fluctuations´ source further away from the 2DEG.

Magnetotransport characterization was carried out on 4 by 4 mm$^2$ square-shaped samples cleaved from the wafer center; eight Indium contacts were positioned at the corners and the midpoints of the sides. Sample preparation and characterization at 1.3 K (electron density, mobility and the presence/absence of parallel conductance) was performed in a $^4$He cryostat. Lowest-temperature measurements were performed in a $^3$He/$^4$He toploading dilution refrigerator system with a base temperature of ~10 mK, using an AC current in the nA range and standard Lock-in techniques. As R$_{xx}$ is expected to vary as exp(-Δ/2k$_B$T), the activation energy Δ could be determined via temperature dependent measurements. When discussing the results, we concentrate on the values for $\Delta_{5/2}$, but results for $\Delta_{7/3}$ and partially for $\Delta_{8/3}$ are also noted in table 1.

**Table 1.** Sample overview. "dop type" designates the applied doping scheme, d$_{eff}$ is the mean setback distance, μ and n the electron mobility and density measured at 1.3 K and Δ 5/2, Δ 7/3 and Δ 8/3 the determined gap energies at the respective filling factors.

| Sample | μ (cm$^2$/Vs) | n (cm$^{-2}$) | $\Delta_{5/2}$ (10$^{-3}$ e$^2$/4πεl$_B$) | $\Delta_{5/2}$ (mK) | $\Delta_{7/3}$ (mK) | $\Delta_{8/3}$ (mK) |
|---|---|---|---|---|---|---|
| A | 1,91·10$^7$ | 2.84·10$^{11}$ | 0,44 | 48 | 56 | 24 |
| B | 1,99·10$^7$ | 2.61·10$^{11}$ | 0,52 | 54 | 66 | 12 |
| C | 1,92·10$^7$ | 2.95·10$^{11}$ | 0,25 | 28 | 26 | - |
| D | 2,10·10$^7$ | 2.54·10$^{11}$ | 0,84 | 86 | 100 | 49 |
| E | 1,92·10$^7$ | 2.13·10$^{11}$ | 1,43 | 135 | 125 | - |
| F | 2,07·10$^7$ | 2.58·10$^{11}$ | 0,55 | 57 | 76 | 23 |
| G | 1,31·10$^7$ | 2.46·10$^{11}$ | 0,03 | 3 | 5 | - |
| H | 1,55·10$^7$ | 2.66·10$^{11}$ | 0,31 | 33 | 19 | - |
| I | 1,96·10$^7$ | 2.87·10$^{11}$ | 0,47 | 51 | 34 | 11 |

## 3. Results

In figure 2(a), the Arrhenius plots (ln(R$_{XX}$) plotted versus 1/T) for samples A to I are shown, along with the best linear fits to the data (red lines) from which the noted gap energies are derived. Figure 2(b), (c) and (d) exemplarily depict the temperature-dependent resistances for sample E ($\Delta_{5/2}$ 135 mK), sample I ($\Delta_{5/2}$ 51 mK) and sample H ($\Delta_{5/2}$ 33 mK), respectively. Please note that the electron densities of these nine samples differ within a small margin - except for sample E, having despite its very low electron density of 2.13·10$^{11}$ cm$^{-2}$ a high gap energy.

To allow for comparing samples with different densities we will use furtheron the normalized gap energies in units of 10$^{-3}$ times the Coulomb energy $E_{Coul} = e^2 / 4\pi \cdot \varepsilon \cdot l_B$, where ε = 12.9 is the dielectric constant for GaAs and $l_B = \sqrt{\hbar / e \cdot B_{5/2}}$ is the magnetic length (see table 1).

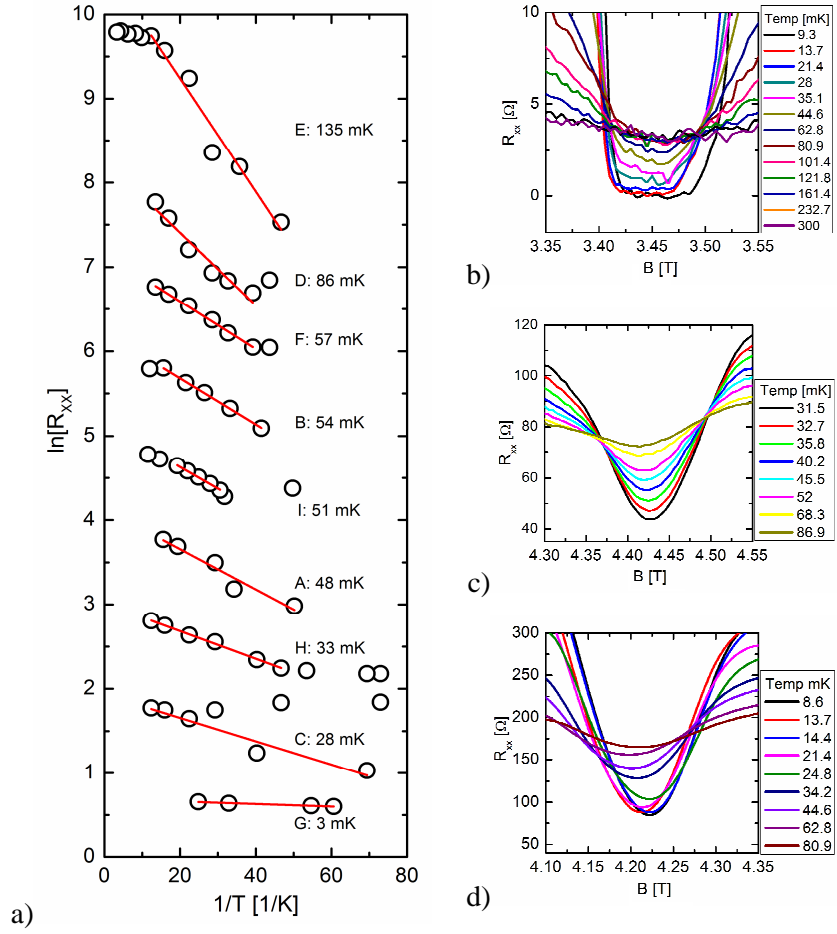

**Figure 2.** (a) Arrhenius plots for samples A to I from which the noted gap energies were calculated (in descending order). The measurements have arranged offsets for clarity, red lines represent the best linear fits to the data. (b), (c) and (d) are close-ups of the temperature-dependent $R_{XX}$-traces in the magnetic field range corresponding to filling factor 5/2. Shown are the temperature-dependencies of (b) sample E, (c) sample I and (d) sample H with normalized gap energies of 1.43, 0.47 and 0.31 (in units of $10^{-3}\ E_{Coul}$).

Figure 3 compares our normalized gap energies to other results from literature. The gap values are plotted versus electron density in units of the Coulomb energy. One has to point out that there are many gap energies reported that are considerably higher than ours. However, the colour code reveals the preparations prior to the respective measurements: Blue indicates a gated and red an illuminated structure. The black markers show samples that were measured in "as grown" condition. To our knowledge, the only 5/2 gap energy measurements performed in such a sample state were reported by Miller et al. [10] (black diamond) and by Gamez and Muraki [21] (black crossed squares), both of them reporting lower gap energies than our best sample.

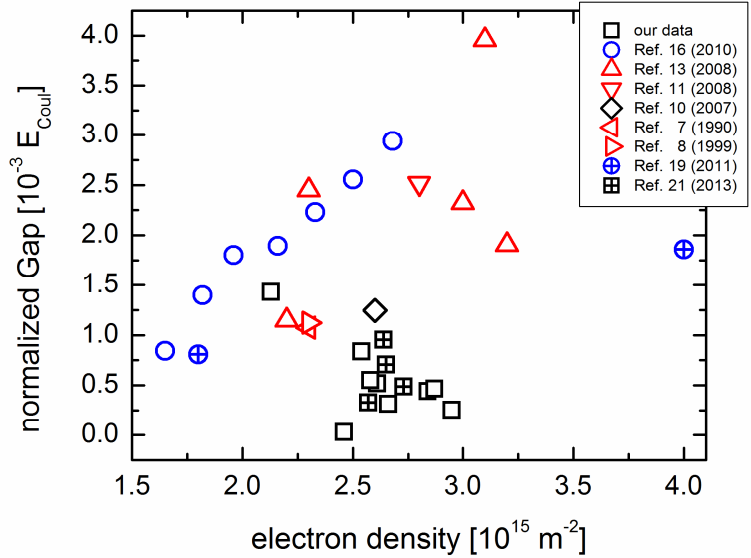

**Figure 3.** Comparison of our $\Delta_{5/2}$ data to literature. The colour code indicates the preparations prior to measurement (Blue: gates, red: illumination, black: "as grown" state).

In order to ensure the high quality of the samples used in this investigation, sample E was measured in magnetic field range corresponding to filling factors 2 to 3. Distinct minima were observed at 11/5, 21/9, 7/3, 22/9, 5/2, 23/9, 8/3, 19/7 and 14/5, although for 21/9, 22/9 and 23/9 no corresponding Hall plateau was found. However, the finding of the 21/9 FQHS would be the first ever reported to our knowledge, so this $R_{XX}$ minimum might as well have a different, yet unknown to us, origin. These findings, along with three well developed re-entrant states, confirm not only the high sample quality, but also show that sample E is on par with those reported e.g. in [11] and [13], despite those samples having considerably higher electron mobilities and 5/2 gap energies.

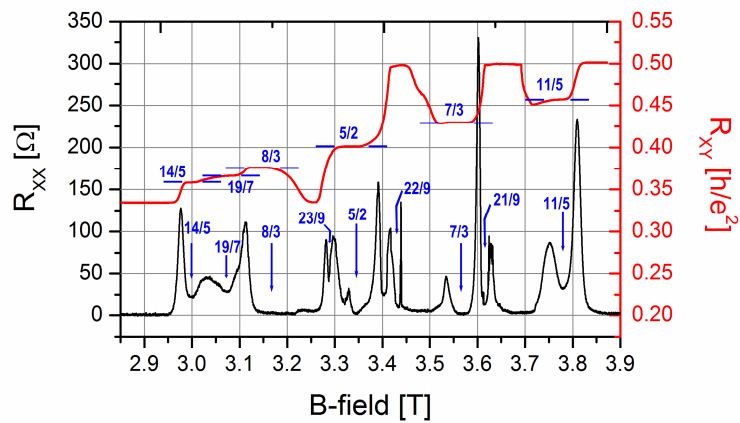

**Figure 4.** Magnetotransport measurement of sample E in the magnetic field range corresponding to $\nu = 3$ to $\nu = 2$. In the Hall traces, represented by the red line, plateaus for 14/5, 19/7, 8/3, 5/2, 7/3 and 11/5, along with three very pronounced re-entrant states are observed. $R_{XX}$ additionally features minima for 23/9, 22/9 and 21/9, although without clearly visible corresponding Hall plateaus.

A set of careful measurements with low currents and at temperatures below 15 mK on a variety of samples (samples A, E, G, H and two additional structures of the same type) with different gap energies suggest a qualitative relation between gap size and Hall features, especially with reference to appearance and development of the (currently known) up to four re-entrant states between filling factors 2 and 3 (figure 5).

Growth parameters and thus, the corresponding sample quality, are subject to variations over time – be it over the course of months like e.g. vacuum quality or even during a single growth run, like growth rates or substrate temperature. So, for a meaningful analysis, one has to ensure that reproducibility of sample structure and 2DEG characteristics on the highest level. First, we compare sample A and B, which were both produced under nominally identical growth conditions with a time lag of ten months.

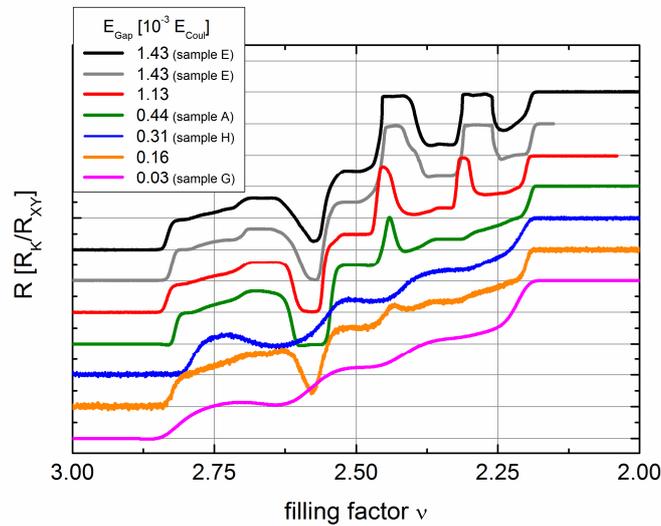

**Figure 5.** Hall traces of 2DEGs with different gap energies. The y-axis covers the Hall resistance values with arranged offsets for clarity. The black and grey lines represent two different measurements of the same sample.

Sample B being the newer structure shows a gap energy of 0.52, close to the 0.44 acquired from the older sample A. Noting the higher electron mobility of sample B, this modest increase in gap energy may be assigned to the slow thermal self-cleaning of the source materials observed during every growth campaign [30]. On a shorter timescale of several weeks however, we were able to reproduce sample characteristics within a margin of ± 2% for electron density and ± 3% for electron mobility. In order to evaluate the effect of growth parameter variations on the 5/2 gap energy however, a characterization of electron mobility is not sufficient, as mentioned above.

The most critical scattering process - long range Coulomb interaction between 2DEG electrons and ionized donors – can be minimized by increasing the distance between dopants and the 2DEG. For samples C and D, as well as samples A/B and E, only $d_{eff}$ was changed from 72.5 nm (C, A/B) to 102.5 nm (D and E) for both doping layers. According to theory, the RI scattering rate scales with $1/d^{2.5}$ [37], leading to a reduction of RI scattering when increasing the distance between 2DEG and ionized donors. As RI-scattering is considered to be the dominant process limiting gap energies in high quality 2DEGs, an increase of the gap energy can be expected. Although there was no significant change in electron mobility (see table 1), $\Delta_{5/2}$ increased by a factor of more than three from 0.25 (sample C) to 0.84 (sample D) and by 2.75 from 0.52 (sample B) to 1.43 (sample E) respectively. Sample C was grown one year earlier, thus, its gap energy can be extrapolated to 0.30, which leads to a factor of 2.8.

In addition, these findings allow us to evaluate the screening effect of heavy X-band electrons in AlAs (samples A, B and E are QW-doped) versus a conventional DX-doping (samples C and D). A comparison of the gap energies indicates an increase by more than 70 percent due to X-electron screening: From 0.25 (sample C) to 0.44 (sample A) and from 0.84 (sample D) to 1.43 (sample E).

Please note that all samples have comparable electron mobilities of approximately $2 \cdot 10^7$ cm$^2$/Vs despite the gap energies varying overall by a factor of five – strongly supporting earlier investigations by Umansky et al. [31] that RI scattering, despite only weakly affecting the electron mobilities, is in fact the main obstacle on the road to higher 5/2 gap energies.

However, the effects of background impurities on sample quality still need to be taken into account. Background impurities are commonly held responsible for over 80% of all mobility affecting scattering events even in 2DEG samples of ultra high quality [30]. In order to lower the amount of background impurities, we reduced the substrate temperature during growth operation from 650°C (sample F) to 630°C (sample D). In this way impurity segregation in the growing structure as well as the amount of impurities evaporating from the resistive substrate heating can be reduced. Although the effect on the electron mobility was negligible ($2.1 \cdot 10^7$ cm$^2$/Vs for both samples), the value of the 5/2 activation gap increased from 0.55 (sample F) to 0.84 (sample D).

Another approach to reduce the amount of background impurities is to reduce the temperature of the aluminium source while the quantum well is grown. The temperature was gradually reduced during growth of the lower setback region and raised thereafter in the upper setback, thus creating setback layers with a gradually lowering Al-fraction (from 0.25 to a final value of 0.16) in the direction of the quantum well (see figure 6). The technique has the additional advantage of reducing the amount of highly reactive aluminium near (and segregation into) the quantum well. In addition to a clear increase in the electron mobility – $1.31 \cdot 10^7$ cm$^2$/Vs for sample G vs. $1.55 \cdot 10^7$ cm$^2$/Vs for sample H – also an increase in the activation energy by a full order of magnitude from 0.03 to 0.31 was found.

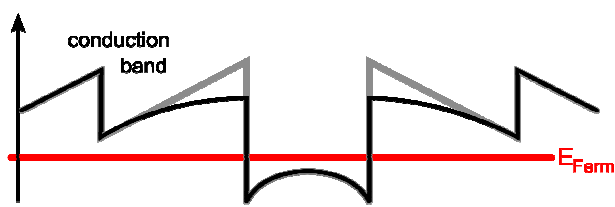

**Figure 6.** Schematic drawing of the conduction bands of sample G with constant (grey) and sample H with gradually reduced Al-fraction (black). The Fermi level is indicated in red.

The lower electron mobilities of these samples however indicate that the primary factor limiting mobility (and gap energy) is not the aluminium-caused BI scattering. Samples A-C and E all have the "high" aluminium fraction of 25% and still show substantially higher electron mobilities. So, only part of the increase in gap energy can be attributed to the reduced Al-fraction. To get a deeper insight in the effect, a further decrease of the aluminium fraction would be desirable. However, the effective growth rate was already well below 0.2 µm/hour for AlAs, and at such low values ensuring a stable growth rate is challenging. Additionally, reducing the growth rates always comes at the cost of increased incorporation of residual impurity atoms originating from the background that would affect sample quality.

Finally we investigated sample I, where the over-doping factor γ for the lower doping layer was set to 1.5. For the upper doping a γ of 1.2 was chosen – a reduction of 30% with respect to sample B, were γ is 1.6 for both doping layers. In this way, we expect the structure to be (even) more suitable for all kinds of experiments requiring top gates. Despite the reduced over-doping and hence reduced RI-screening, $\Delta_{5/2}$ suffers only a very moderate reduction from 0.52 (for sample B) to 0.47.

## 4. Conclusions and Outlook

In summary, our findings clearly confirm that electron mobility alone is not the relevant quantity to judge the quality of a 2DEG structure with respect to FQHS features and especially the gap energy of the 5/2 state. However, a comparative analysis of $R_{XX}$ and $R_{XY}$, measured at very low temperatures, allows for a qualitative prediction of gap values. Admittedly, the quality of these features are extremely temperature-dependent (shown e.g. in [32, 38]), so this kind of characterization would require a setup with good reproducibility in terms of sample temperature in the low mK range.

As for growth parameters, a substantial increase in gap energies was found when implementing the QW-doping instead of the standard DX-doping scheme. One has to point out that DX-doped samples may yield superior gap energy values – compared to QW-doped ones – however, after illumination. Also a higher amount of over-doping in QW-doped samples leads to more pronounced SdH features [31] and supposedly to higher gap energies as well. The accompanying hysteresis (and necessity for high bias gate voltages) however forbids this otherwise preferable technique for a variety of applications. We found an increase in gap energy by more than a factor of two via increasing the setback distance, as expected. It would be interesting to explore the limits of this parameter and determine the "ideal" setback, where the positive effect of higher distance is not offset by the inevitable decrease in electron density. The investigations on reduced background impurities support the idea that a reduced level of background impurities will do good not only to electron mobility [39], but also to gap energy.

We hope that combining our results in one growth recipe to obtain 2DEG structures with the stacked benefits of all individual optimizations may provide 5/2 gap energies in as-grown samples that are on par with illuminated or gated ones.

## 5. Acknowledgements

We gratefully acknowledge the financial support of the Swiss National Foundation (Schweizerischer Nationalfonds, NCCR "Quantum Science and Technology") and the BMBF (Bundesministerium für Bildung und Forschung), Nr 01BM900.

**Appendix**

Figure A.1 shows characterization data of five single-interface 2DEG samples (not shown in table 1) produced with an identical set of growth parameters. As can be seen, the electron density can be reproduced within ± 2%, whereas the normalized electron mobility per density (confined to small ranges, electron mobility scales in good approximation linearly with density) is stable within ± 3%. We would like to point out that these data were obtained without illuminating the samples. In this "dark" state DX-doped single interface 2DEGs are known to be highly sensitive to structural changes.

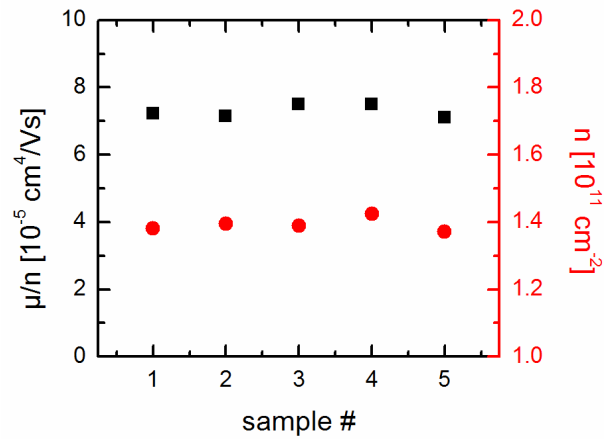

**Figure A.1.** Reproducibility of identically grown single-interface 2DEG-structures with DX-doping.

**Table A.** List of growth parameters varied and investigated. Shown is 5/2 gap energy $\Delta_{5/2}$ in units of the Coulomb energy; electron density and mobility at 1.3 Kelvin; the effective thickness of the setback layers; total depth of the 2DEG below the surface; amount of dopants in the upper and lower doping layer (according to calibration obtained from thick uniformly doped GaAs); ratio between upper and lower doping, chosen to yield a symmetric electron wave function centered in the quantum well (with the exception of sample I); amount of over-doping γ where applicable; doping scheme; aluminium content in the setback close to the quantum well; growth temperature; and "age", i.e. in which month of the growth campaign it was produced. Please note that storage time proved to be uncritical, as our samples do not show notable differences in µ, n and SdH features (measured at 300 mK), even with as much as two years between two experiments.

| Sample | A 0728B | B 0523A | C 0712D | D 0702A | E 0427C | F 0705D | G 0111A | H 0113C | I 0724A |
|---|---|---|---|---|---|---|---|---|---|
| $\Delta_{5/2}$ [$10^{-3} E_{Coul}$] | 0.44 | 0.52 | 0.25 | 0.84 | 1.43 | 0.55 | 0.03 | 0.31 | 0.47 |
| n [$10^{11}$ cm$^{-2}$] | 2.84 | 2.61 | 2.95 | 2.54 | 2.13 | 2.58 | 2.46 | 2.66 | 2.87 |
| µ [$10^{7}$ cm$^{2}$/Vs] | 1.91 | 1.99 | 1.92 | 2.10 | 1.92 | 2.07 | 1.31 | 1.55 | 1.96 |
| $d_{eff}$ [nm] | 72.5 | 72.5 | 72.5 | 102.5 | 102.5 | 102.5 | 102.5 | 102.5 | 77.5 |
| Depth [nm] | 225 | 225 | 225 | 200 | 250 | 200 | 200 | 200 | 195 |
| $dop_{up}$ [cm$^{-2}$] | 25·10$^{11}$ | 35·10$^{11}$ | 22·10$^{11}$ | 42·10$^{11}$ | 31·10$^{11}$ | 42·10$^{11}$ | 35·10$^{11}$ | 35·10$^{11}$ | 24·10$^{11}$ |
| $dop_{low}$ [cm$^{-2}$] | 7·10$^{11}$ | 10·10$^{11}$ | 6.5·10$^{11}$ | 8.5·10$^{11}$ | 9·10$^{11}$ | 8.5·10$^{11}$ | 7·10$^{11}$ | 7·10$^{11}$ | 9·10$^{11}$ |
| $dop_{up}$ / $dop_{low}$ | 3.5 | 3.5 | 3.5 | 5 | 3.5 | 5 | 5 | 5 | 2.7 |
| γ | 1.5 | 1.6 | 1 | 1 | 1.4 | 1 | 1 | 1 | 1.2/1.5 |
| dop type | QW | QW | DX | DX | QW | DX | DX | DX | QW |
| Al-fraction | 0.25 | 0.25 | 0.25 | 0.25 | 0.25 | 0.25 | 0.25 | 0.16 | 0.25 |
| Tgrowth [°C] | 630 | 630 | 630 | 630 | 630 | 650 | 630 | 630 | 630 |
| "age" | 8 | 18 | 7 | 19 | 17 | 19 | 13 | 13 | 20 |